# Machine Learning Assisted Long-Range Wireless Power Transfer

Likai Wang[1], Yuqian Wang[1], Shengyu Hu[1], Yunhui Li[2], Hong Chen[1], Ce Wang[1*], and Zhiwei Guo[1*]

[1]MOE Key Laboratory of Advanced Micro-Structured Materials, School of Physics Science and Engineering, Tongji University, Shanghai 200092, China
[2]Department of Electrical Engineering, Tongji University, Shanghai 201804, China

[*] Correspondence to: 21153@tongji.edu.cn (C.W.); 2014guozhiwei@tongji.edu.cn (Z.G.)

**Abstract**—Near-field magnetic resonance wireless power transfer (WPT) technology has garnered significant attention due to its broad application prospects in medical implants, electric vehicles, and robotics. Addressing the challenges faced by traditional WPT systems in frequency optimization and sensitivity to environmental disturbances, this study innovatively applies the gradient descent optimization algorithm to enhance a system with topological characteristics. Experimental results demonstrate that the machine learning-optimized Su-Schrieffer-Heeger (SSH)-like chain exhibits exceptional performance in transfer efficiency and system robustness. This achievement integrates non-Hermitian physics, topological physics, and machine learning, opening up new avenues and showcasing immense potential for the development of high-performance near-field wave functional devices.

**INTRODUCTION**

Magnetic resonance wireless power transfer (WPT) technology [1,2] has shown great potential in various fields such as smartphones, robotics, medical implants, and electric vehicles, attracting extensive research interest [3-8]. However, in standard second-order resonant systems, the coupling strength between the transmitter and receiver decreases sharply as the transmission distance increases, leading to a significant drop in long-distance transfer efficiency [9-12]. To extend the transfer distance without sacrificing efficiency, researchers have proposed adding multiple relay coils between the transmitter and receiver, forming a uniform chain structure similar to dominoes [13-17]. However, this domino chain structure still faces numerous challenges in achieving stable and efficient long-distance WPT [18, 19]: (1) When the spacing between relay coils is too small, strong near-field coupling effects can cause significant frequency splitting, which is influenced by the spacing and number of coils, making it difficult to achieve stable WPT at a fixed operating frequency [20, 21]; (2) When the spacing between relay coils is larger, the frequency splitting caused by near-field coupling decreases, and the operating frequency can be fixed at the resonant frequency of the coils, but the transfer efficiency decreases accordingly, making it highly challenging to balance stability and efficiency in long-distance WPT; (3) In traditional domino systems [13], the magnetic field is almost uniformly distributed across all coils, resulting in a significant increase in overall energy loss (Ohmic loss); (4) As the number of relay coils increases, structural errors gradually accumulate, leading to a sharp decrease in transfer efficiency and increased fluctuations. Therefore, there is an urgent need to develop innovative technologies or methods to optimize existing technologies and achieve stable and efficient long-range WPT [22-26].

In recent years, the rapid development of artificial intelligence has greatly promoted the widespread application of machine learning algorithms in the field of physical multi-parameter optimization, like photonic structure design [27-29], evaporative cooling experiment optimization [30-32], and quantum state preparation [33-37]. In these application scenarios, the relationship between optimization parameters and target quantities is often regarded as a "black box", and neural network structures driven by reinforcement learning or active learning are commonly used to optimize target parameters. In particular, several recent studies have demonstrated the successful integration of machine learning techniques in electromagnetic WPT system optimization [38-41]. However, in specific situations, there is a clear functional relationship between parameters and target quantities, which provides the possibility for applying optimization strategies such as gradient descent or self-feedback [42].

In this work, we propose an innovative WPT scheme based on gradient descent optimization algorithm (GDOA), aimed at achieving optimal transfer efficiency. We conducted an in-depth analysis of how the coupling strength between two nearest neighbor resonant coils affects transfer efficiency, and cleverly introduced GDOA for optimization. Through GDOA, we can accurately determine the specific position distribution of the resonator corresponding to the optimal transfer efficiency. To verify the effectiveness of the GDOA, we considered a second-order model with known optimal distribution exact solutions. We found that the solution provided by the GDOA is highly consistent with the exact solution and have conducted experimental verification (see Supplementary Information A [43]). Furthermore, we apply the GDOA to a high-order chain model consisting of ten oscillators. The GDOA successfully predicted the optimal model and achieved perfect impedance matching at the target frequency. Compared with traditional models such as uniform chain and SSH chain models, our optimization demonstrates significant advantages and potential in performance. This result fully demonstrates the outstanding performance of GDOA in achieving optimal transfer efficiency.

**RESULTS**

*Model for long-range WPT system.-* Take the classical one-dimensional domino chain WPT system as an example, the dynamics at driven frequency can be described by the coupled mode theory (CMT) [44]:

$$i\frac{d\vec{a}}{dt} = H\vec{a} + e^{-i\omega t}\sqrt{2\gamma_t}\vec{s}$$
$$H = \sum_i [(\omega_0 - i\Gamma - i\gamma_i)c_i^\dagger c_i + \kappa_i c_{i+1}^\dagger c_i] + H.c. \quad (1)$$

where $c_i^\dagger (c_i)$ is the creation (annihilation) operator of the $i$-th resonator located at position $r_i$. For simplification, the resonant frequency $\omega_0$ and dissipative loss $\Gamma$ of all coils is identical, and the radiative loss $\gamma_i$ is nonzero only for the transmitter $\gamma_1 = \gamma_t$ and the receiver $\gamma_L = \gamma_r$. The nearest-neighbor coupling strength between the $i$-th and $i+1$-th coils is represented by $\kappa_i = e^{-(r_{i+1}-r_i)/d_0}$, with the normalized distance constant $d_0 = 0.0181$. $\vec{a} = (a_1, a_2, ..., a_L)^T$ is the vector representation for the all the complex field $a_l$ on the $l$-th coil, $\vec{s} = (1, 0, ..., 0)^T$ represents the source driving on the transmitter at strength of $\gamma_t$. A straightforward calculation shows that the transmission ($\tau$) can be given by:

$$\tau(\omega) = \left|\sqrt{2\gamma_t} a_L\right| = \left|2\sqrt{\gamma_t \gamma_r} \left(\frac{1}{H - \omega I}\right)_{L,1}\right|, \quad (2)$$

The objective function is to find the optimal $\{\kappa_l\}$ reproducing the largest $\tau(\omega)$ by adjusting the position distribution $\{r_l\}$.

*GDOA*.-To be compact, we directly optimize $\{\kappa_l\}$ which is equivalent to the optimization of $\{r_l\}$. We introduce the Green's function matrix as

$$G(\omega) = (H - \omega I)^{-1}. \quad (3)$$

According to Eq. (2), the transition rate is related to G as $T = \sqrt{2\gamma_t \gamma_r}|G_{L,1}|$. Taking the derivative of both sides of the equation $G(H - \omega I) = I$ with respect to $\kappa_l$, we obtain

$$\frac{\partial G}{\partial \kappa_l} = -G \frac{\partial H}{\partial \kappa_l} G, \quad (4)$$

With Eq. (4), it is easy to show that

$$\frac{\partial G_{L,1}}{\partial \kappa_l} = -G_{L,l} G_{l+1,1} - G_{L,l+1} G_{l,1} \equiv \delta G_l, \quad (5)$$

as long as the real part of $\delta G_l / G_{L,1}$ is positive, an increase in the coupling $\kappa_l$ can lead to a larger transition rate. Together with the constraint of $\Pi_{l=1}^{L-1} \kappa_l = T_p$, we can then introduce a gradient flow $\kappa \to \kappa + r\Delta\kappa$ to optimize $T$ with

$$\Delta\kappa = \text{Re}\left(\frac{\delta G}{G_{L,1}}\right) - T(\kappa)\left[\text{Re}\left(\frac{\delta G}{G_{L,1}}\right) \cdot T(\kappa)\right], \quad (6)$$

where $\mathbf{T}(\kappa) \propto (1/\kappa_1, ..., 1/\kappa_{L-1})$ is a unit vector. Following Eq. (6), we can then optimize $T$ by repeating the gradient descent. To ensure physical consistency during optimization, it is essential to maintain the invariance of the product of coupling strengths. Therefore, we introduce a modified gradient update rule that removes components that would violate this constraint. A detailed mathematical derivation and explanation of this constraint-preserving mechanism are provided in see Supplementary Information B [43]).

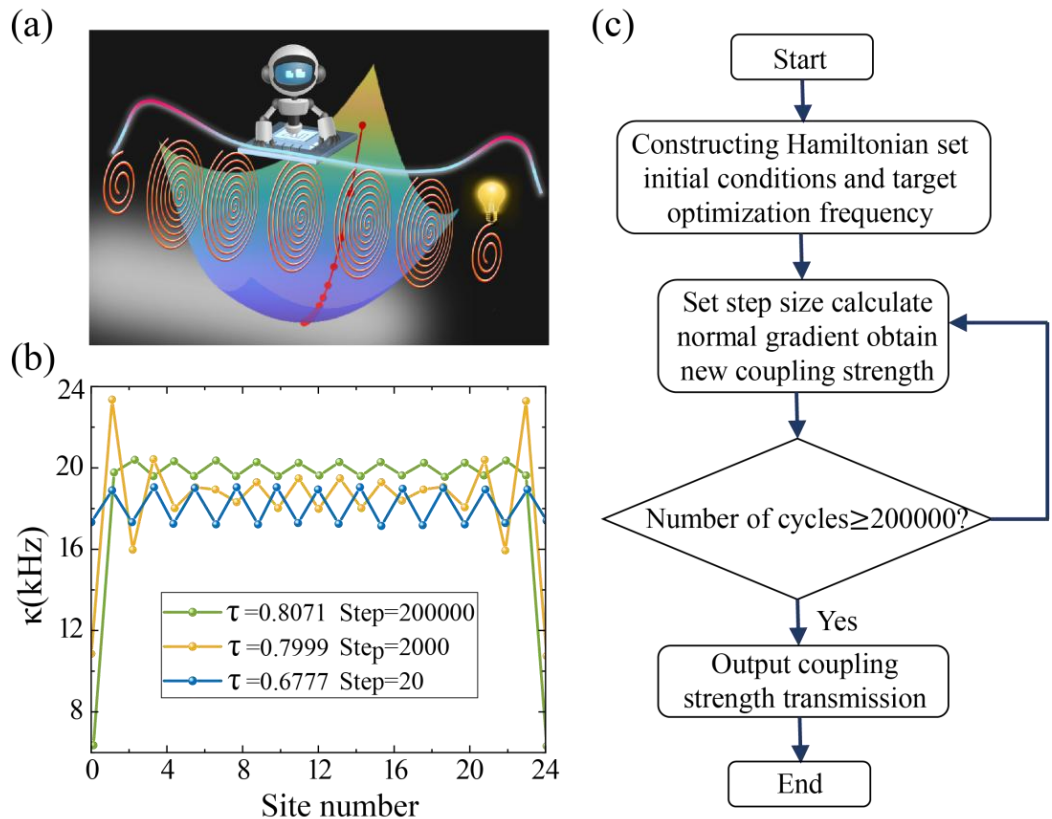

**Figure 1**. **GDOA for long-range domino-like WPT system**. (a) Conceptual diagram of the GDOA model. (b) Changes in efficiency and coupling distribution at training steps of 20, 2000, and 200000. The initial configuration is set to a uniform distribution, and at step 20, the GDOA extracts the SSH model. The optimized result is an improved version of the SSH model, with its coupling strengths on both sides exhibiting different characteristics compared to the traditional SSH structure. (c) Flowchart of machine learning based GDOA.

Figure. 1(b) presents three scenarios in the training process: step sizes of 20, 2000, and 200000, when $\omega=1$ kHz, $\gamma_0=0.2$ kHz, $\gamma_t = \gamma_r = 2$ kHz, and $L=24$. Specifically, we investigate a domino-like system consisting of $L=10$ uniformly arranged resonant units, which is used as the initial step for the GDOA. Following the flowchart shown in Fig. 1(c), iterative optimization is performed at a specific optimization operation frequency. Given the similarity between the gradient optimization chain and the SSH chain, we have chosen the SSH chain as one of the comparison models, which is based on the structural characteristics observed under very small number ($\leq 20$) of iterations. To further demonstrate the advantages of the GDOA over traditional optimization approaches, we compared it with commonly used methods such as Bayesian optimization and Monte Carlo simulations. Results show that GDOA achieves significantly better convergence speed and scalability. A comprehensive comparison including computational efficiency, dimensional adaptability, and accuracy is presented in see Supplementary Information C [43]).

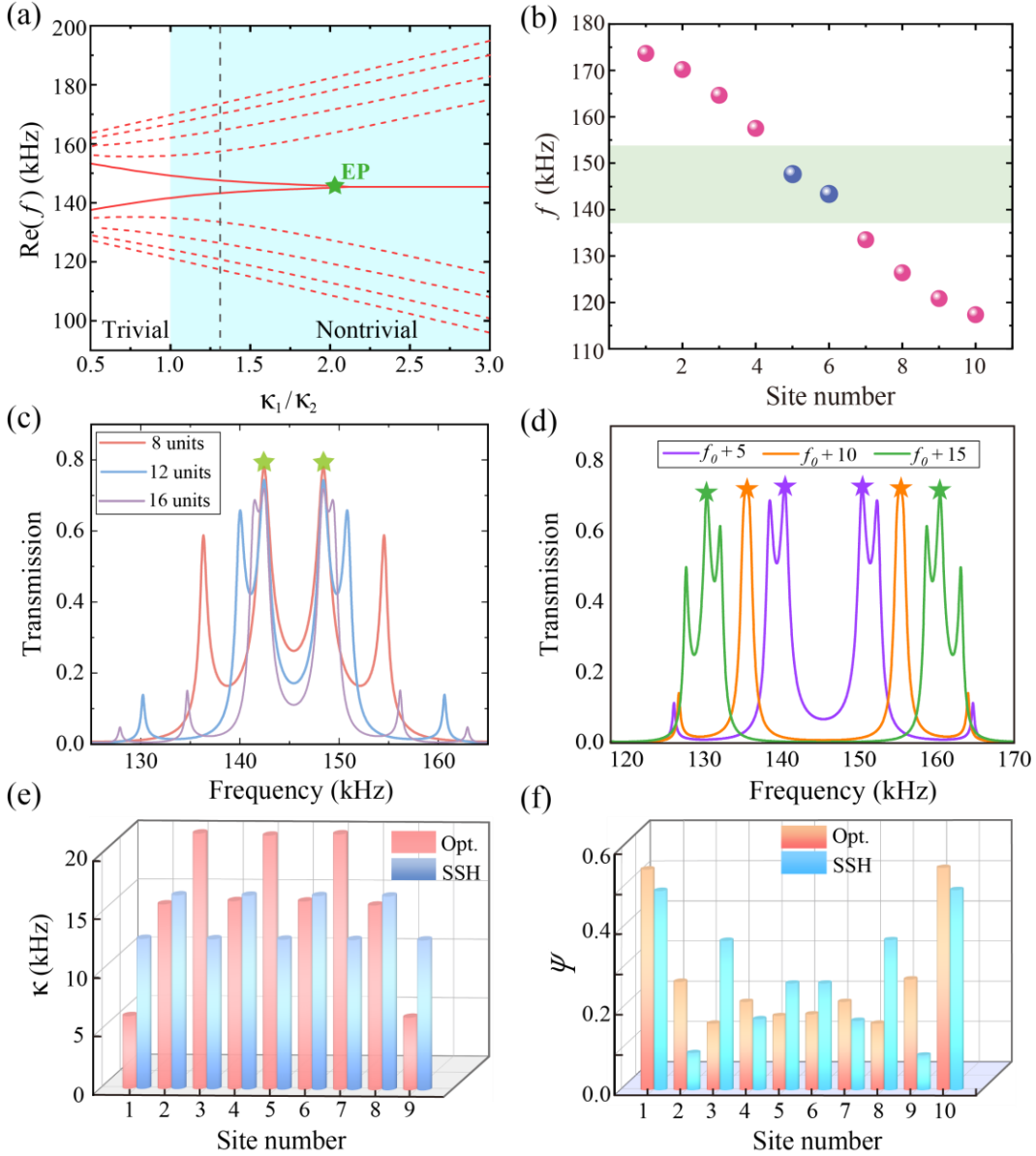

**Figure 2**. **The comparison between GDOA results and SSH chains.** (a) Real part of the eigenfrequency, with blue and white areas representing the topologically nontrivial phase and the trivial phase, respectively. The red solid and dashed lines indicate edge modes and bulk modes, respectively. The black dashed line marks the experimentally selected $\kappa_1/\kappa_2 = 1.296$. (b) Eigenvalue spectrum of the topological nontrivial SSH chain, with topological modes located within the bandgap (green area). (c) Gradient optimization results for different numbers of oscillator units. (d) Gradient optimization results for different target frequencies. (e) Distribution of coupling strength with respect to the position of resonant coils, with red bars representing the gradient optimized chain and blue bars representing the SSH chain. (f) Distribution of wave functions $\Psi$, with the same legend as in (e).

At the 30$^{th}$ iteration of the optimization process, the system's dynamic behavior demonstrates the characteristic features of the SSH model. This model, recognized for its topological robustness, ensures the stability of energy transfer by virtue of its topologically non-trivial phase. To establish a rigorous comparative research framework, the SSH model identified during the optimization process is employed as the reference benchmark for subsequent optimization analysis. For the SSH-like WPT chain, these characteristics arise from the alternating coupling strengths between adjacent resonators, which create a synthetic one-dimensional lattice with a bandgap [24]. The topology of this bandgap determines whether edge states—localized modes at the chain's ends—emerge. These states are intrinsically robust to disorder, as their existence is tied to the global symmetry of the system rather than specific local parameters.

Figure. 2(a) illustrates the real part of the eigenfrequency as a function of the ratio of the strong coupling area to the weak coupling area with $\kappa_1/\kappa_2$ ( Here we fix $\kappa_1$ to be constant at 12.78 and then change the strength of $\kappa_2$ ). Notably, $\kappa_1/\kappa_2=1$ serves as the critical point distinguishing between topologically trivial and nontrivial states. When $\kappa_1/\kappa_2<1$, a trivial bandgap opens near the frequency $\omega_0$; whereas when $\kappa_1/\kappa_2>1$, a topologically nontrivial bandgap opens around $\omega_0$, accompanied by the emergence of two topological edge states. Specifically, at the exceptional point $\kappa_1/\kappa_2=2.05$, the eigenfrequencies and eigenmodes of these two edge states coalesce. In the exact phase of parity-time-symmetry $(\kappa_1/\kappa_2<2.05)$, the eigenfrequencies of the two edge states are purely real but deviate from $\omega_0$. However, once entering the broken phase of parity-time-symmetry $(\kappa_1/\kappa_2>2.05)$, the eigenfrequencies of these two edge states become complex, with the fixed real parts $\omega_0$ (as indicated by the red solid line). It is worth noting that the remaining eight eigenfrequencies of the bulk states remain purely real throughout the phase transition (red dashed line). For details on the imaginary parts of the eigenfrequencies, please refer to the Supplementary Information D [43]. In the experimental section, we selected $\kappa_1/\kappa_2=1.296$ (the black dashed line), with corresponding coupling strengths and distances of $\kappa_1 = 16.58$ kHz, $d_1 = 2.5$ cm, and $\kappa_2 = 12.79$ kHz, $d_2 = 3.4$ cm. The corresponding eigenvalue distribution is shown in Fig. 2(b), where the blue dots specifically highlight the unique topological modes within the bandgap.

Furthermore, we have presented the optimization results under different conditions: Fig. 2(c) vividly illustrates the optimization scenario for the number of resonators, while Fig. 2(d) clearly demonstrates the optimization results for three different target frequencies. This fully demonstrates that our proposed novel WPT scheme can flexibly meet the optimization requirements under various objectives. Additionally, figure. 2(e) compares the coupling coefficient $\kappa$ relationship between the optimized chain (red) and the SSH chain (blue). It is evident that there are significant differences in coupling strength between the optimized chain and the SSH chain, and the coupling strength of the optimized chain exhibits a symmetric distribution around the structural center (site number is 5). By calculating the wave function $\Psi$ for both chains, we found (as shown in Fig. 2(f)) that although the intensities of the 10 resonators are similar, the wave function of the optimized chain is better localized at the ends of the structure, while the dissipation in the middle is also reduced. This indicates that, compared to the SSH chain, the optimized chain may exhibit superior robustness.

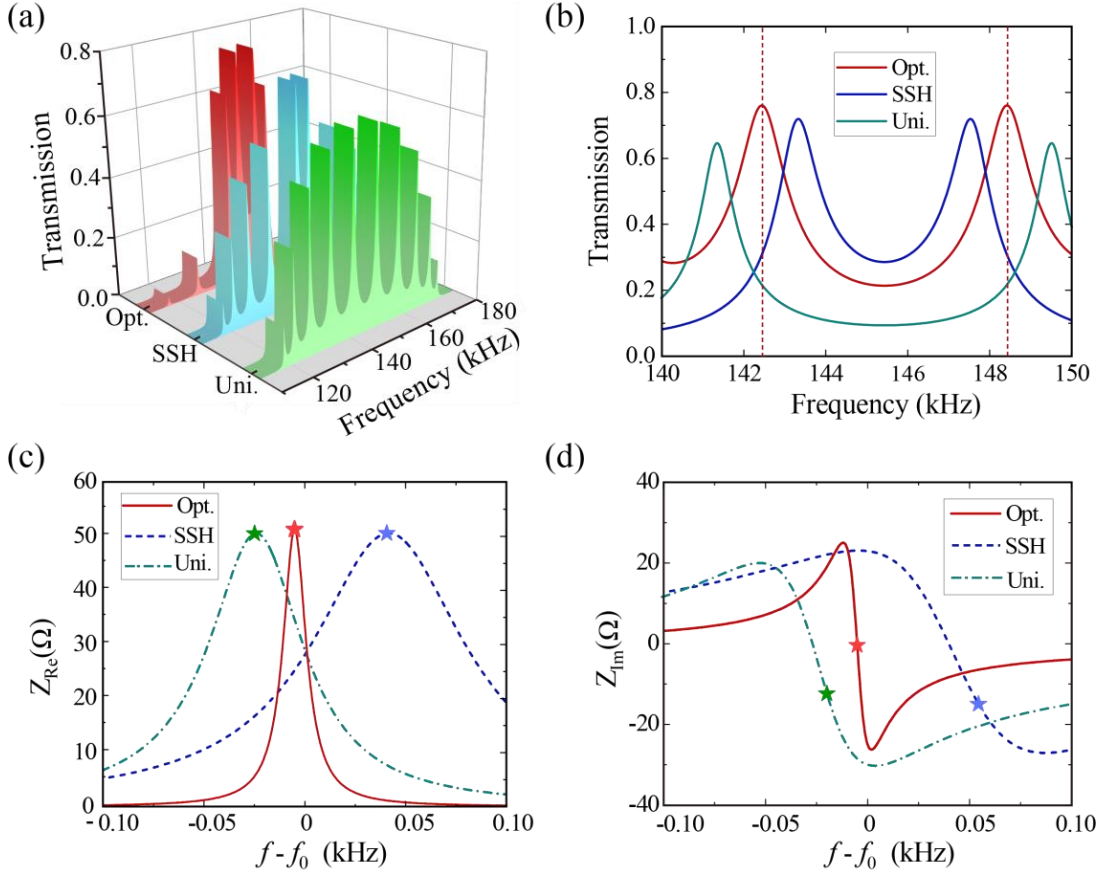

**Figure 3**. **Calculated transmission and impedance of three chains.** (a) Theoretical calculations have been conducted on the relationship between transmission and frequency for optimized chains (red), SSH chains (blue), and uniform chains (green). (b) The transmission of the three chains near the operating frequency. (c) The real part and (d) imaginary part of their impedances. The pentagrams mark the frequency corresponding to the highest transmission.

We designed three types of chains, with their coupling strengths adjusted by varying the distances. The specific parameters are as supplementary material [43]. With other parameters kept constant, it is evident from the transmission rate versus frequency relationship shown in Fig. 3(a) that the optimized chain (red) exhibits a significantly higher transmission rate compared to the SSH chain (blue) and the uniform chain (green). Figure. 3(b) provides a local magnification of Fig. 3(a), focusing on the optimized frequencies around approximately 148.4 kHz (and 142.4 kHz). Due to the symmetry of the system structure, a high transmission point naturally emerges at 142.4 kHz, adjacent to the primary point at 148.4 kHz. Notably, we calculated and compared the frequency values corresponding to the highest transmission rates for each chain: approximately 142.4 kHz for the optimized chain, approximately 141.3 kHz for the SSH chain, and approximately 143.3 kHz for the uniform chain.

Next, we delve into the impedance matching analysis of these three types of chains from the perspective of circuit theory. Initially, we set the resistance on the load side to 50Ω as a benchmark. Subsequently, we calculate the mapped impedance $Z_{ref}$ and the input impedance $Z_{in}$. In this process, the closer the input impedance is to the ideal state (the real part of the resistor is 50Ω, and the imaginary part of the reactive impedance is 0Ω), it indicates better impedance matching, which subsequently leads to a significant enhancement in the system's transfer efficiency. Based on Kirchhoff's laws:

$$\begin{aligned}
& I_{10}(i\omega L_{10} + \frac{1}{i\omega C_{10}}) - i\omega M_9 I_9 + I_{10} R_S = 0; \\
& Z_{ref9} = \frac{-i\omega M_9 I_{10}}{I_9}; M_9 = \frac{2L_9 \kappa_9}{\omega}; \\
& Z_{ref8} = \frac{-i\omega M_8 I_9}{I_8}; M_8 = \frac{2L_8 \kappa_8}{\omega}; \\
& ... \\
& Z_{ref1} = \frac{-i\omega M_1 I_2}{I_1}; M_1 = \frac{2L_1 \kappa_1}{\omega}; \\
& Z_{in1} = Z_{ref1} + \frac{1}{i\omega C_1} + i\omega L_1;
\end{aligned} \qquad (7)$$

where $C_1 = C_2 = ... = C_{10}$, $L_1 = L_2 = ... = L_{10}$.

The real and imaginary parts of the input impedance for the three types of chains are presented in Figs. 3(c) and 3(d). The red solid line, blue dashed line, and green dashed line depict the impedance characteristics of the optimized chain, SSH chain, and uniform chain, respectively. The pentagram markers indicate the positions with the highest transmission rates on each line. It is evident from the figures that the optimized chain exhibits the best impedance matching effect, followed by the SSH chain, while the uniform chain shows the poorest matching effect. This clear trend is consistent with the calculation results based on CMT presented in Figs. 3(a) and 3(b).

We have successfully verified the theoretical transmission results through experiments. The experimental setup, where two non-resonant coils (Tx and Rx) are connected to the two ports of the vector network analyzer (VNA, SIGLENT SNA5084X), is shown in Fig. 4(a). In this experiment, we employed three types of one-dimensional dimer chains, each consisting of 10 identical sub-wavelength resonant coils carefully constructed. These coils were uniformly wound on circular acrylic plates and equipped with a load capacitance of 5.15 nF. The coils had a diameter of 12 cm, featuring a double-layer winding design with a total of 20 turns, and the inductance value was stable at approximately 222.5 µH, with a resonant frequency close to 145.4 kHz. The non-resonant coils serving as transmitters and receivers were not equipped with load capacitances, had the same diameter of 12 cm, a single-layer winding of 20 turns, and inductance value of 110 µH. The photos of the enlarged resonant coil (right) and non-resonant coil (left) are shown in Fig. 4(b). Notably, we ensured that the total length of the three chains (the distance from the first resonant coil to the tenth resonant coil) was consistently 27 cm, guaranteeing the same transmission distance across all three chains.

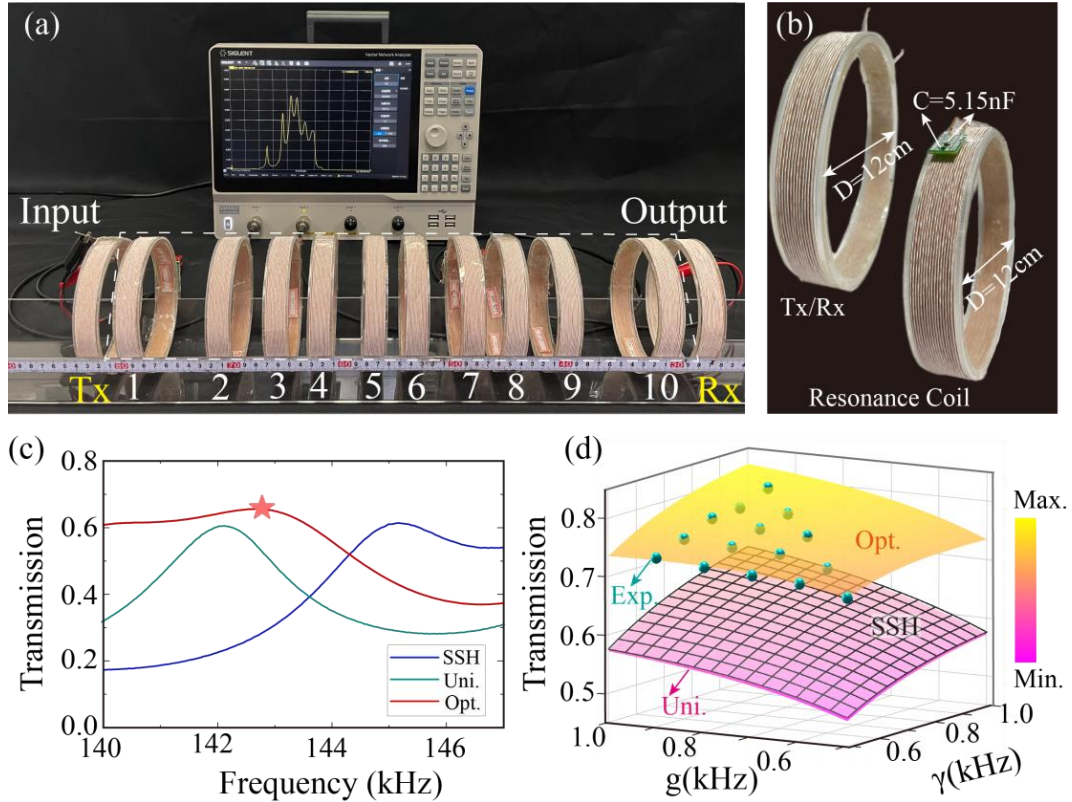

**Figure 4**. **Experimental comparison of transmission and robustness of three types of chains.** (a) The photo of the experimental setup consists of 10 resonant coils, a left non-resonant source coil, a right non-resonant receiver coil, and a VNA. (b) Photos of the enlarged resonant coil (right) and non-resonant coil (left). (c) Experimental measured transmission of the three chains, where the pentagram represents the maximum transmission of the optimized chain at the operating frequency. (d) The relationship diagram between transmission, gain, and loss rate of three types of chains. The green spheres correspond to the experimental results.

Utilizing VNA, we connected two non-resonant coils to its ports to assess the transmission of three distinct chains. The experimental outcomes are presented in Fig. 4(c), showcasing the transmission of the optimized, uniform, and SSH chains, which are in agreement with our theoretical predictions (detailed experimental spectral data can be found in the Supplementary Information E [43]). Additionally, figure. 4(d) unveils the transmission trends of these three chains across varying gain $g$ and loss $\gamma$. Notably, across the entire test range, the optimized chain (experimental data marked with green dots) exhibits significantly better transmission performance compared to the SSH chain and uniform chain. The scalability and generality of the GDOA approach are validated through additional experiments on extended systems with 12 and 14 resonators, as detailed in the Supplementary

Information F [43]. In all configurations, GDOA consistently achieves high transmittance (>0.6) at the target frequency of 142.4 kHz, showing minimal sensitivity to chain length variations. These results confirm the method's robustness and adaptability across different system scales, effectively addressing limitations of fixed-size assumptions. This scalability underscores the algorithm's strong potential for practical, flexible, and efficient long-range WPT applications.

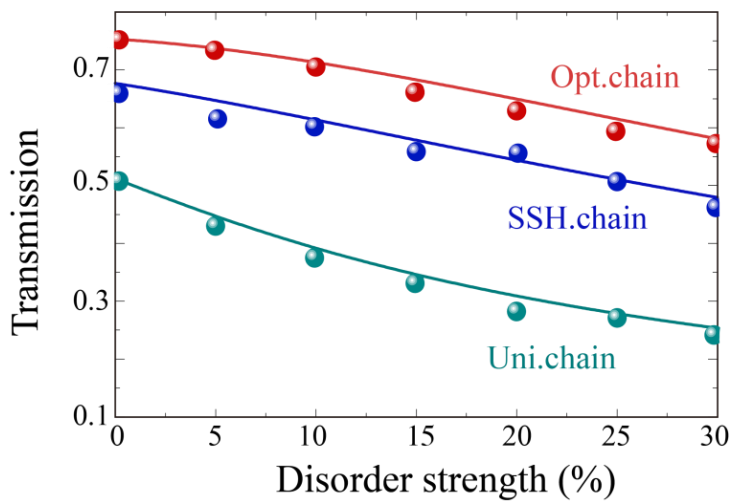

**Figure 5**. **The transmission of three types of chains varies with the intensity of disorder disturbance.** The calculated and measured results are marked by the solid lines and dots, respectively.

In finite topological systems, zero modes, as topological edge states, often deviate from the exact zero-energy state in finite Hermitian lattices due to coupling between these edge states. This coupling inevitably weakens topological protection and reduces the robustness of SSH chains [45, 46]. Nevertheless, these localized modes are more stable than the normal bulk states [45, 46]. Therefore, we further compared the robustness of the three chains. The magnitude of disorder perturbations was determined by the product of the perturbation intensity and a random number ranging from 0 to 1, with disorder introduced by adjusting the coupling coefficients. We further experimentally investigated the robustness of these three chains under different perturbation intensities. We display a comparison of the transmission rates of the three chains under different perturbation intensities. Specifically, figure 5 illustrates the comparison results at a perturbation intensity of 30%. The theoretical calculations here were based on averages from 10000 random perturbation instances as a reference benchmark. (Detailed data can be found in the Supplementary Information E [43]). The experimental results clearly demonstrate that the optimized chain not only exhibits higher transmission efficiency, but also inherits the robustness of the SSH model to structural perturbation immunity. The robustness of the optimized chain stems from its topological characteristics. In finite SSH-like systems, edge states are protected against perturbations that preserve the system's symmetry. Finally, we compared the system parameters and performance of our work with those reported in other studies, as summarized in Table 1. To provide a more intuitive representation of the performance differences among various systems, we use letter grades: A+ denotes the best performance, A represents excellent performance, A- indicates slightly weaker performance, and so on. As shown in the table, our system demonstrates the highest overall robustness.

**Table .1 Comparison of parameters and performance with other literature**

| Ref. | Number of coils | The resonance order of the system | Coil radius (cm) | Distance to diameter ratio | Transmission range | Transmission capability (efficiency) | Limited at resonance frequency | Robust to structural fluctuations | Robust to transmission distance | Overall robustness |
|---|---|---|---|---|---|---|---|---|---|---|
| [2] | 2 | 2 | 30 | 8 | Mid-far | B (0.4) | No | No | No | B |
| [20] | 2 | 2 | 29 | 2.1 | Mid-far | A (0.9) | No | No | Yes | A |
| [12] | 2 | 3 | 4.3 | 4 | Mid-far | A (0.9) | Yes | No | Yes | B |
| [17] | 3 | 3 | 15 | 2 | Mid-far | A- (0.8) | Yes | No | Yes | B |
| [13] | 11 | 11 | 15.5 | 11.6 | Far | A-(0.5) | No | No | No | B- |
| [24] | 10 | 10 | 4 | 10.5 | Far | A (0.6) | Yes | Yes | No | A |
| [23] | 16 | 16 | 2.6 | 9.6 | Far | A-(0.1) | Yes | Yes | No | A |
| [25] | 16 | 16 | 2.6 | 9..6 | Far | B-(0.1) | No | Yes | No | A |
| This work | 10 | 10 | 6 | 11.8 | Far | A (0.64) | No | Yes | Yes | A+ |

## CONCLUSION

In summary, we have optimized the long-range multi-relay coil WPT system using the GDOA, innovatively proposing a novel WPT scheme that is both efficient and robust. Through dual verification by CMT and circuit theory, we have proven that the optimized transmission significantly surpasses that of SSH chains and uniform chains, which has been further confirmed through experiments. Furthermore, we have conducted a thorough analysis of the stability performance of uniform chains, SSH chains, and optimized chains at their respective optimal operating frequencies. The results indicate that the optimized chain system exhibits the strongest anti-interference capability against perturbations in position-dependent coupling strength, implying higher stability when faced with disturbances such as transmission distance variations. In industrial and daily applications, there is an urgent demand for long-distance, high-stability, and efficient chained WPT technology, particularly in areas such as the long arms of mobile machinery, high-voltage power transmission detection, and feedback devices [47, 48]. Machine learning not only helps to accurately identify topological phases [49, 50], but also significantly improves the technical performance of one-dimensional systems. Considering that most power supplies in real-life scenarios have fixed output frequencies, we have optimized the chained WPT using the GDOA, enabling it to achieve high transfer efficiency and demonstrate strong adaptability to changes in power supply output frequency. Meanwhile, in the supplementary materials, we have detailed the exceptional flexibility of the GDOA in optimizing frequency and the number of oscillators, which is highly attractive for practical industrial applications.

## ACKNOWLEDGMENT

This work is supported by the National Key R&D Program of China (Nos. 2023YFA1407600 and 2021YFA1400602), the National Natural Science Foundation of China (Nos. 12104342, 12374294, and 52477014), the Interdisciplinary key project of Tongji University (No. 2023-1-ZD-02), the Shanghai Science and Technology Commission Project (No. 2021SHZDZX0100), and the Chenguang Program of Shanghai (No. 21CGA22).

## AUTHOR DECLARATIONS

### Conflict of Interest

The authors have no conflicts to disclose.

## AUTHOR CONTRIBUTIONS

L. Wang and Y. Wang contribute equally to this work.

**Likai Wang:** Data curation (equal); Investigation (equal); Validation (equal); Writing – original draft (equal). **Yuqian Wang:** Data curation (equal); Investigation (equal); Validation (equal); Writing – original draft (equal). **Yunhui Li:** Funding acquisition (equal); Formal analysis (equal); Writing – review & editing (equal). **Shengyu Hu:** Investigation (equal); Writing – review & editing (equal). **Hong Chen:** Conceptualization (equal); Investigation (equal); Writing – review & editing (equal). **Ce Wang**: Conceptualization (equal); Funding acquisition (equal); Formal analysis (equal); Investigation (equal); Writing – review & editing (equal). **Zhiwei Guo**: Conceptualization (equal); Funding acquisition (equal); Formal analysis (equal); Investigation (equal); Writing – review & editing (equal).

## DATA AVAILABILITY

The data that support the findings of this study are available from the corresponding author upon reasonable request.

## CODE AVAILABILITY

All the codes that support the findings of this study are available from the corresponding authors upon reasonable request.